\documentclass[a4paper,11pt]{article}
\pdfoutput=1 

\usepackage{jheppub} 
\usepackage[T1]{fontenc} 
\usepackage{amsmath}

\def\l{\left(}
\def\r{\right)}

\title{\boldmath $K^{+}\rightarrow \mu^{+}\nu_{\mu} \nu \bar \nu$
and $K^{+}\rightarrow e^{+}\nu_{e} \nu \bar \nu$
decays within the Chiral Perturbation Theory}

\author{D. Gorbunov}
\author{and A. Mitrofanov}
\affiliation{Institute for Nuclear Research of the
Russian Academy of Sciences, Moscow 117312, Russia \\and\\
Moscow Institute of Physics and Technology, Dolgoprudny 141700, Russia}

\emailAdd{gorby@ms2.inr.ac.ru}
\emailAdd{mitrofanov@phystech.edu}

\abstract{Decays $K^{+} \rightarrow \mu^{+}\nu_{\mu} \nu \bar \nu$ and
  $K^{+}\rightarrow e^{+}\nu_{e} \nu \bar \nu$
are examined to the leading order in momenta in the framework of
Chiral perturbation theory. Predictions of the Standard Model for
the muon and electron differential energy spectra and branching ratios
of $K_{\mu 3 \nu}$ and $K_{e 3 \nu}$ are presented.}

\usepackage{subfig}
\usepackage{graphicx}	
\graphicspath{{/}}	
\DeclareGraphicsExtensions{.pdf,.png,.jpg}	
\begin{document}

\maketitle
\flushbottom

\section{Introduction}

 Standard Model of particle physics (SM) allows at tree level of
 perturbation theory fully leptonic 4-body kaon decays $K^{+}
 \rightarrow \mu^{+} \nu_{\mu} \bar \nu_l \nu_l $ and $K^{+}
 \rightarrow e^{+} \nu_{\mu} \bar \nu_l \nu_l $, where
 $l=e,\mu,\tau$. However they occur only in the fourth order of weak
 coupling constant $g$ (that is in the second order of the Fermi
 constant $G_F$) and hence are expected to be extremely rare.  We
 calculate the amplitudes of these processes exploiting the Chiral
 Perturbation Theory (ChPT)\,\cite{Gasser:1982ap}. In this paper we
 perform calculations of the process amplitudes to the leading order
 in momenta expansion, $\mathcal{O}(p^2)$, obtain the charged lepton
 spectra and estimate the corresponding branching ratios.

Experimental searches for the kaon leptonic 4-body decays, starting
from Refs.\,\cite{Cable:1972wm,Pang:1989ut} reveal no evidence for the
processes so far, which is not surprising given the smallness of
estimated SM branching ratios $\sim 10^{-16}$\,\cite{Pang:1989ut}. The
kaon experiments with large statistics, e.g. E494, NA62 (with about
$10^{13}$ charged kaons), can either find a hint of new physics or
place new much stronger limits on the partial widths $K_{\mu3\nu}$ and
$K_{e3\nu}$. So the correct SM estimate is needed to tell one case
from another. Ultimately, in the absence of new physics signal, an
experiment  with sufficiently large statistics would allow to probe the
new ChPT form factors related to the neutral weak boson exchange.

In practice, only a part of the phase space of final
particles is available for experimental study. In our case
the only one particle in the final state, the charged lepton, is observable,
hence one has to calculate the charged lepton spectra in $K_{\mu3\nu}$
and $K_{e3\nu}$.

The theoretical predictions for the charged lepton spectra are also
required, as $K_{\mu3\nu}$ and $K_{e3\nu}$ contribute to the
background in searches for new particles in kaon decays, $K^+\to\mu^++
X$, $K^+\to\mu^++ X+Y$, $K^+\to\mu^++ X+Y+Z$, etc where neutral
particles, $X$, $Y$, $Z$, etc (including hypothetical ones) are
invisible. Well-known examples are searches for sterile neutrinos,
$K^+\to\mu^++ N$, see e.g. Ref.\,\cite{Artamonov:2014urb} for recent
experimental results and Ref.\,\cite{Gorbunov:2007ak} for possible
motivation, searches for hidden photons, $K^+\to \mu + \nu +A$, see
e.g. Ref.\,\cite{Beranek:2012ey} for recent results and
Ref.\,\cite{Pospelov:2007mp} for possible motivation.  Though the
contribution of $K_{\mu 3 \nu}$ is beyond the immediate registration,
in many model of new physics we have no definite predictions for
the signal strength. Extensive searches for new physics may,
sooner or later, bring to the situation where detailed predictions of
the SM contributions to $K_{l 3 \nu}$ are required to either further
constrain or study the new physics models.

Processes $K_{\mu3\nu}$ and $K_{e3\nu}$ have been considered
previously either within a hypothetical models of strong neutrino
interactions \cite{Bardin:1970wq,Pang:1989ut} or within the SM where
only  charged current contributions of electron and muon doublets have
been accounted for \cite{Pang:1989ut}. Neutral currents
have been considered in Ref.\,\cite{Beranek:2012ey}. In the present
paper we calculate all the relevant SM leading order contributions
working within ChPT.

The paper is organized as follows. In Sec.\,\ref{Sec:lagr} we derive
the interaction lagrangian relevant for the kaon leptonic 4-body
decays, present the Feynman diagrams of the processes and
corresponding amplitudes. The numerical results for lepton spectra and
partial decay widths are presented in Sec.\,\ref{sec:results}. We
conclude in Sec.\,\ref{sec:conclusions}.
Appendix \ref{squared} contains the details of
calculation of squared matrix elements of $K_{\mu3\nu}$ and
$K_{e3\nu}$, integration over the 4-body phase space is performed in
appendix \ref{phase-space}.

  \section{Lagrangian and diagrams}
\label{Sec:lagr}
\begin{figure}[!htb]
   \center{\includegraphics[width=\textwidth]{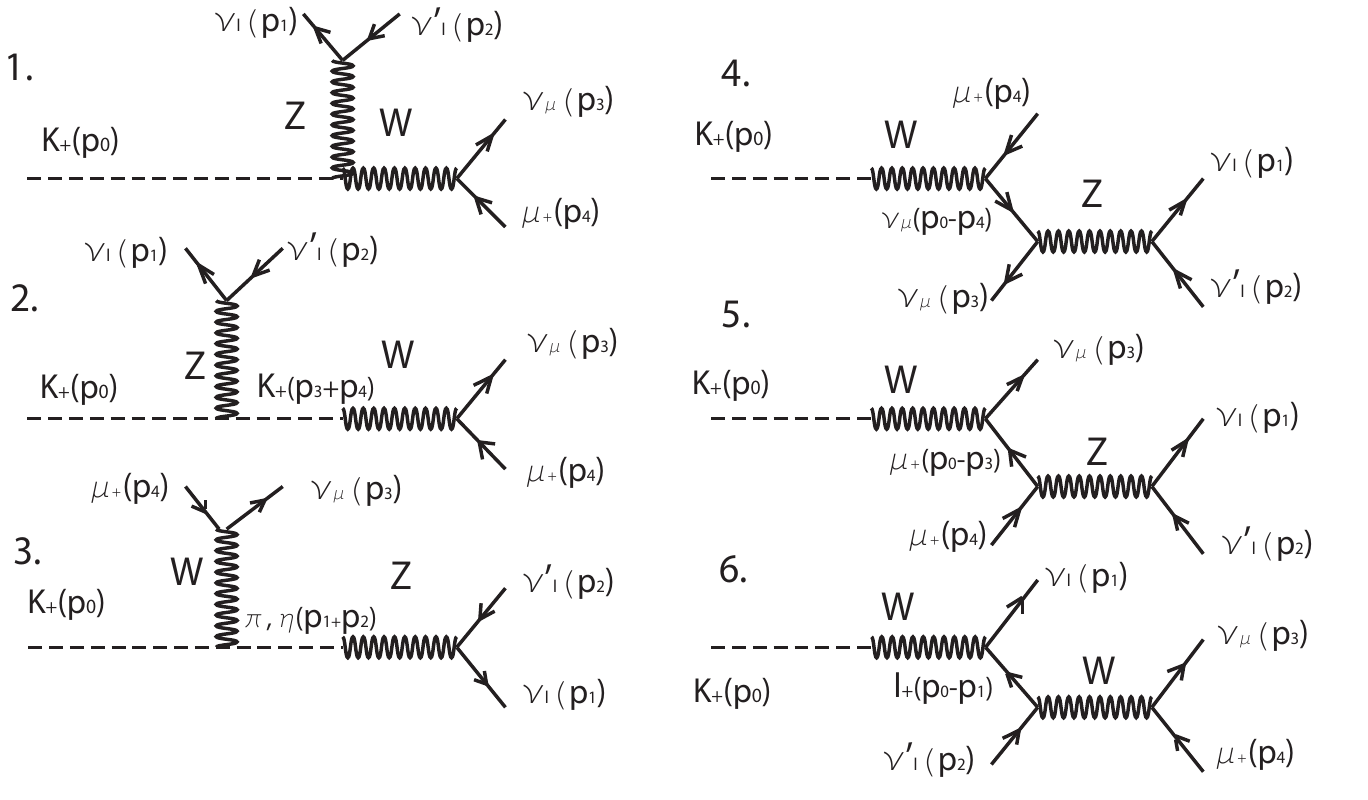}}
	 \caption{\label{diagrams}
Feynman diagrams of the process
$K^{+} \rightarrow \mu^{+} \nu_{\mu} \bar \nu_l \nu_l $. Similar
           diagrams contribute to the
decay $K^{+} \rightarrow e^{+} \nu_{e} \bar \nu_l \nu_l $.}
\end{figure}

 We are using the standard ChPT\,\cite{Gasser:1982ap}, which properly
 describes the dynamics of the octet of light pseudoscalar mesons
forming the unitary matrix
\begin{equation}
\label{field}
U = {\rm e}^{i\,\sqrt{2}\,\frac{\Phi}{F}}\,,
\;\;\;\;
\Phi=
 \l
\begin{matrix}
\frac{\pi^0}{\sqrt{2}} + \frac{\eta}{\sqrt{6}} & \pi^+ & K^+ \\
\pi^- & -\frac{\pi^0}{\sqrt{2}} + \frac{\eta}{\sqrt{6}} & K^0 \\
K^- & \bar K^0 & -\frac{2\,\eta}{\sqrt{6}}  \\
\end{matrix}
\r\,,
\end{equation}
$F$ is meson decay constant, which numerical value is specified in
due course.
To the leading order in momentum the model lagrangian can be obtained
from\,\cite{Bijnens:1994qh}
 \begin{equation}
\label{lagr}
 L_2 = \dfrac{F^2}{4} \text{Tr}\left[ D_{\mu} U D^{\mu} U^{\dag} + \chi
 U^{\dag} + \chi^{\dag} U \right] ,
 \end{equation}
 where
\begin{equation}
\label{def}
D_{\mu}U = \partial_\mu U - i r_{\mu} U + i U l_{\mu}\;,\;\;\;\; \chi
= 2B\times \text{diag}\l m_u,\,m_d,\,m_s\r\,.
\end{equation}
Here $l_{\mu}$ and $r_{\mu}$ are introduced as follows: $r_{\mu} = v_{\mu}
+ a_{\mu}$ and $l_{\mu} = v_{\mu} - a_{\mu}$, where $v_{\mu},a_{\mu}$
are sums of the weak and the electromagnetic fields,
interacting with vector and
axial quark currents in SM. These interactions for the relevant set of
three light quarks, $q\equiv (u,\,d,\,s)^T$, which masses enter $\chi$
in \eqref{def}, reads
\begin{equation}
L_{\text{int}}
= \bar q \gamma^{\mu} (v_{\mu} + \gamma_5 a_{\mu}) q\,.
\end{equation}
SM gives the following quark couplings to
 charged $W_\mu^\pm$ and neutral $Z_\mu$ weak gauge bosons:
\begin{equation}
L_{SM(u,d,s)} = -\dfrac{g}{2 \sqrt{2}} \left[ W_{\mu}^{+} \bar q  T_{+}
\gamma^{\mu} (1 - \gamma_5) q + h.c. \right] - \dfrac{g}{2 \cos
  \theta_W}Z_{\mu} \bar q \gamma^{\mu} (v_{nc} -\gamma_5 a_{nc} ) q
\;,
\end{equation}
where we introduce
\[
T_{+} =  \l
\begin{smallmatrix}
0 & V_{ud} & V_{us} \\ 0 & 0 & 0 \\ 0 & 0 & 0  \\ \end{smallmatrix}
\r\,,
\]
with $V_{ud}$ and $V_{us}$ standing for the corresponding elements of the
Cabibbo--Kobayashi--Maskawa mixing matrix,
and
\begin{equation*}
v_{nc}=\text{diag}
\left(\frac{1}{2}-\frac{4}{3}\sin^2\theta,-\frac{1}{2}+\frac{2}{3}\sin^2\theta,-\frac{1}{2}+\frac{2}{3}\sin^2\theta\right),\;\;\;
a_{nc}=\text{diag} \left(\frac{1}{2},-\frac{1}{2},-\frac{1}{2}\right),
\end{equation*}
with $\theta$ denoting the weak mixing angle, which relates the gauge
boson masses as $M_W=M_Z\cos\theta$.
Thus we obtain for the matrices entering \eqref{def}
\begin{equation}
 r_{\mu} = \frac{\sin^2 \theta}{\cos \theta} g Z_{\mu} Q ,\quad
 l_{\mu} = \frac{g}{\cos\theta} \l - \frac{1}{2}\, A + \sin^2
     \theta\, Q \r Z_\mu - \frac{g}{\sqrt2} (W_{\mu}^{+}T_+
 +h.c.)\;,
 \end{equation}
where $Q=\text{diag}(2/3 , -1/3, -1/3)$, $A=\text{diag} (1, -1,
-1)$. Finally we expand the exponent in \eqref{field}, put it into
\eqref{lagr} and take the relevant for our study part
of ChPT $\mathcal{O}(p^2)$
Lagrangian \eqref{lagr} and leptonic weak current part of the SM
Lagrangian (mass terms are omitted below):
 \begin{equation}\label{rel-lagr}
 \begin{split}
 L &= \frac{i F g^2 \sin^2 \theta }{2 \cos \theta} V_{us} Z^{\mu}
 W^-_{\mu} K^+ \!-\! \frac{F g}{2} V_{us}W^-_{\mu} \partial^{\mu} K^+ \!+\! ig
 \frac{2 \sin^2 \theta - 1}{2 \cos \theta} Z^{\mu} \!\l\partial_{\mu} K^+
 K^- \!-\!  K^+\partial_{\mu} K^-\r  \\ &+ \frac{i g W_{\mu}^- V_{us}}{2
   \sqrt{2}} \left[\partial_{\mu} K^+ \l \frac{\pi^0}{\sqrt 2}
   +3\frac{\eta}{\sqrt 6} \r - K^+ \partial_{\mu} \l
   \frac{\pi^0}{\sqrt 2} +3\frac{\eta}{\sqrt 6} \r\right] - \frac{g F
   Z_{\mu}}{2 \sqrt{2} \cos \theta} \\&\times\partial_{\mu} \l \sqrt{2}
   \pi^0 + \frac{2 \eta}{\sqrt{6}} - \frac{\eta '}{\sqrt 3} \r
 -\frac{g}{2 \sqrt 2} \l W_{\mu}^+ \bar \nu_l \gamma^{\mu} (1
   - \gamma_5) l + h.c \r -\frac{g Z_{\mu}}{4 \cos \theta}
 \bar \nu_l \gamma^{\mu} (1 - \gamma_5) \nu_l
 \end{split}
 \end{equation}
Here $l=e,\,\mu,\,\tau$ and $\nu_l=\nu_e,\,\nu_\mu,\,\nu_\tau$
are charged leptons and corresponding neutrinos and we added
$\eta'$-meson as a natural completion of the set of
relevant mesons.\footnote{Both external fields $a_{\mu},v_{\mu}$ have non-vanishing
traces. It is admissible to gauge not only $SU_V(3) \times
SU_A(3)$ symmetry, but $SU_V(3) \times SU_A(3) \times U_A(1) \times
U_V(1)$, and $\eta'$ emerges in ChPT lagrangian in this case.
Although $U(1)_A$ symmetry is broken due to Chiral anomaly,
Wess--Zumino--Witten term, corresponding to its breaking, is of order
$\mathcal{O}(p^4)$ so we won't take it into consideration. As we find
below, $\eta'$ does not contribute to the processes under discussion
(to the leading order in momenta), so we do
not elaborate on this issue further.}
Diagrams corresponding to
$K^{+} \rightarrow \mu^{+} \nu_{\mu} \bar \nu \nu $ decay
are shown in Fig.\,\ref{diagrams}.
The amplitude of $K_{\mu3\nu}$ can be written as
\begin{equation}
\label{amplitude}
{\cal M}=\frac{FG_F^2V_{us}}{\sqrt{2}} \times \sum_{i=1}^6 M_i \equiv
\frac{FG_F^2V_{us}}{\sqrt{2}} \times M\;,
\end{equation}
where sum goes over all the six diagrams of Fig.\,\ref{diagrams} and
we introduce the Fermi constant, $G_F\equiv g^2/4\sqrt{2}M_W^2$.
The diagrams  contribute to the
amplitude as follows
\begin{align*}
 M_1=&2 \sin^2 \theta \cdot \bar \nu_l (p_1) \gamma^{\lambda}
 (1 - \gamma_5) \nu_{l} (p_2) \cdot \bar \nu_{\mu}(p_3) \gamma_{\lambda} (
 1-\gamma_5 ) \mu(p_4) \\
M_2=&(1-2 \sin^2
 \theta) \cdot \dfrac{(p_3+p_4)^{\lambda} (p_0 + p_3+p_4)^{\rho}}{(p_3+p_4)^2 -
   M_K^2} \cdot \bar \nu_l (p_1) \gamma_{\rho} (1 - \gamma_5) \nu_{l}
 (p_2) \cdot \bar \nu_{\mu}(p_3) \gamma_{\lambda} ( 1-\gamma_5 ) \mu(p_4)
 \\
M_3=& - \frac{1}{2} \cdot \dfrac{(p_1+p_2)^{\lambda} (p_0
   +p_1+p_2)^{\rho}}{(p_1+p_2)^2 - M^2_{\alpha}} \cdot \bar \nu_l (p_1)
 \gamma_{\lambda} (1 - \gamma_5) \nu_{l} (p_2) \cdot \bar \nu_{\mu}(p_3)
 \gamma_{\rho} ( 1-\gamma_5 ) \mu(p_4) \\
M_4=& \frac{1}{2} \cdot
 \dfrac{1}{(p_0-p_4)^2} p^\rho_{0} \cdot  \bar \nu_l (p_1)
 \gamma^{\lambda} (1 - \gamma_5) \nu_{l} (p_2) \cdot \bar \nu_{\mu}(p_3)
 \gamma_{\lambda} ( 1-\gamma_5 ) (\hat p_0 - \hat p_4) \gamma_{\rho}
 (1-\gamma_5) \mu(p_4) \\
M_5=& \frac{1}{2} \cdot p^\lambda_{0}
  \bar \nu_{\mu}(p_3) \gamma_{\lambda} (1 - \gamma_5) \dfrac{ - \hat p_0 +
   \hat p_3 + m_{\mu}}{(p_0-p_3)^2 - m_{\mu}^2} \gamma^{\rho} (4
 \sin^2 \theta - 1 + \gamma_5) \mu(p_4) \cdot \bar \nu_l (p_1)
 \gamma_{\rho} (1 - \gamma_5) \nu_{l} (p_2) \\
M_6=& p^\rho_{0} \cdot \bar \nu_{\mu}(p_3) \gamma_{\lambda} (1-\gamma_5)
 \mu(p_4) \cdot \bar \nu_{l} \gamma^{\lambda} (1-\gamma_5) \dfrac{m_l - \hat
   p_0 + \hat p_1}{(p_0-p_1)^2 - m_l^2} \gamma^{\mu}(1-\gamma_5)
 \nu_{l}(p_2)
 \end{align*}
Similar amplitudes contribute to decay $K_{e3\nu}$ with obvious
replacements  $\mu(p_4)\to e(p_4)$, $\bar\nu_\mu(p_3)\to
\bar\nu_e (p_3)$, and $m_\mu\to m_e$.  In case of decay $K^+
\rightarrow e^+ \nu_e \nu_{\mu} \bar \nu_{\mu}$ diagram $6$ has
resonance divergence associated with muon producing on-shell. We are
dealing it by cutting out the phase space region, corresponding
to $|(p_0-p_1)^2 - m_{\mu}^2| < (30\,\text{MeV})^2$.

\section{Results and discussion}
\label{sec:results}

For the processes under discussion neutrinos can be treated as
massless particles. Then amplitude $M_3$ referring to
diagram 3 in Fig.\,\ref{diagrams}
vanishes ($Z$ mixes with derivative of scalars, see lagrangian
\eqref{rel-lagr} which nullifies the amplitude for outgoing neutrinos
on mass-shell). Also note, that for decays $K^{+} \rightarrow \mu^{+} \nu_{\mu}
\bar \nu_{\mu} \nu_{\mu} $ and $K^{+} \rightarrow e^{+} \nu_{e}
\bar \nu_{e} \nu_{e} $ there are two
identical fermions in the final state and to the amplitudes presented
in Sec.\,\ref{Sec:lagr} must be added the same ones with opposite
signs and changed momenta, $p_1  \leftrightarrow p_3$. In
appendix\,\ref{squared} explicit expressions for the squared amplitudes
are presented. The phase space of the 4-body decays and integration
over the momenta of outgoing fermions are discussed in
appendix\,\ref{phase-space}. In Fig.\,\ref{graf2}
\begin{figure}[!htb]
     \centerline{
     \includegraphics[width=0.5\textwidth]{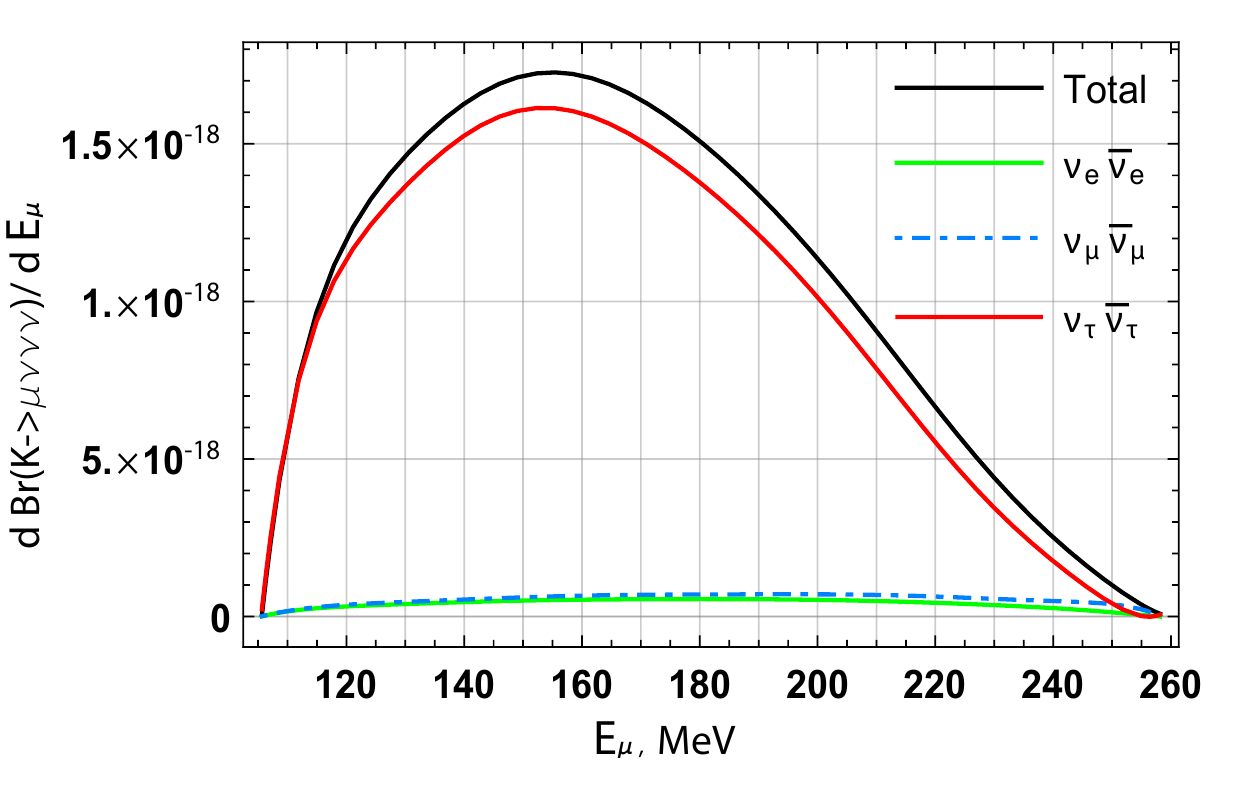}
     \includegraphics[width=0.5\textwidth]{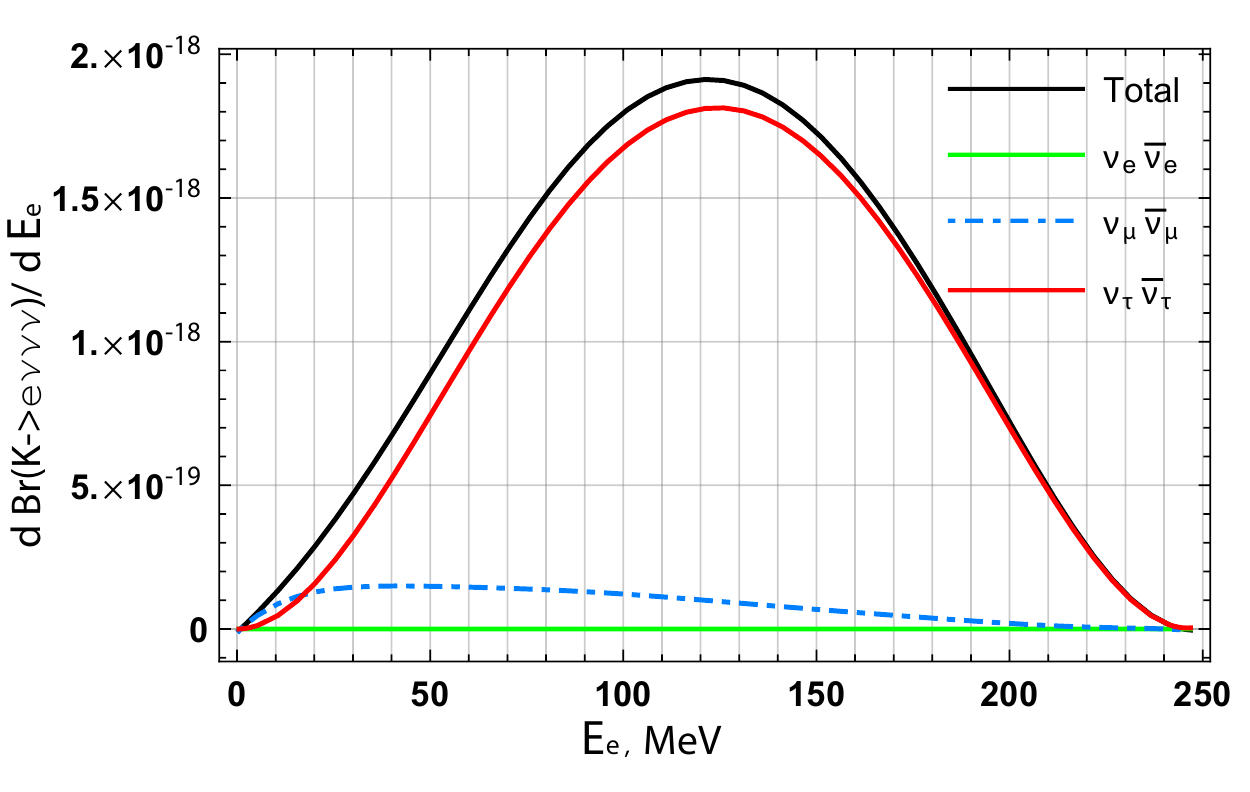}}
     \caption{\label{graf2}Lepton distributions over energy:
Contributions of different final states (electron
  neutrino $\nu_e\bar\nu_e$, muon neutrino $\nu_\mu\bar\nu_\mu$
and tau-neutrino $\nu_\tau\bar\nu_\tau$ pairs)
  to the
differential spectrum of muons from decays $K_{\mu 3\nu}$
{\it (left panel)} and to the
differential spectrum of electrons from decays $K_{e 3\nu}$
{\it (right panel)}, normalized to the corresponding branching ratios.}
\end{figure}
we illustrate contributions of different channels to the differential
spectrum of muons in $K_{\mu 3\nu}$ and electrons in $K_{e 3\nu}$.  As
we can see contributions of the channel with $\tau$-neutrinos
dominate. The difference is associated with the 6th diagram in
Fig.\,\ref{diagrams}, which is suppressed by the $\tau$-lepton
mass. For other channels, the 6th diagram coming with relative
negative sign reduces the sum of others. The details can be traced
with formulas presented in appendix\,\ref{squared}(there the dominant
and relevant for this issue coefficient is denoted as $A$).

\begin{figure}[!htb]
     \centerline
     {\includegraphics[width=0.51\textwidth]{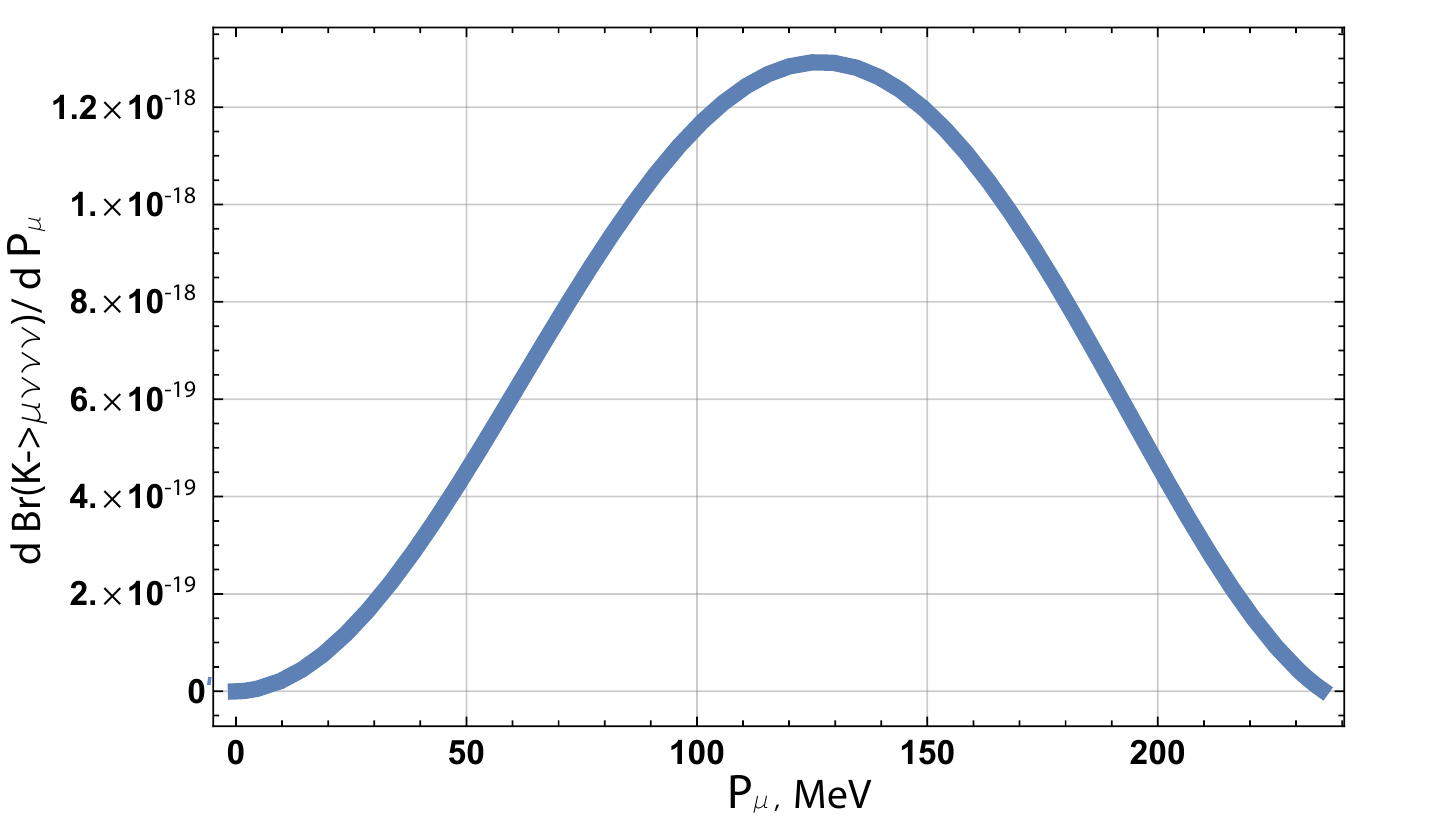}\hskip -0.02\textwidth
      \includegraphics[width=0.51\textwidth]{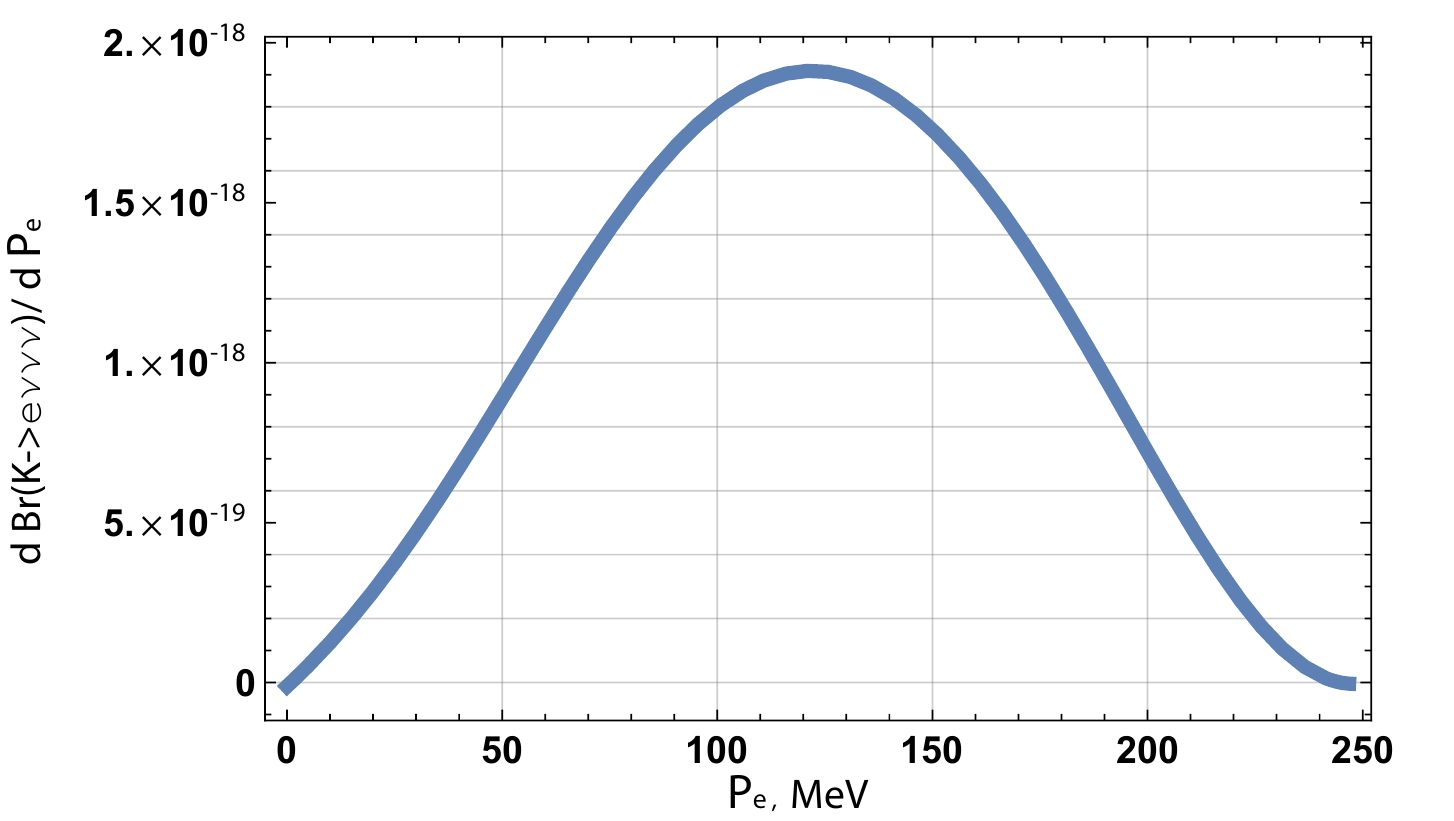}}
     \caption{\label{gra}Momentum distribution of  leptons normalized to the corresponding branching ratios: $K_{\mu     3\nu}$ (left), $K_{e 3\nu}$ (right).}
\end{figure}

The total lepton spectra, sums over all the three final states with
different neutrino flavors, are presented in Fig.\,\ref{graf2} as well. For
experimental application we also show in
Fig.\,\ref{gra}
 the lepton distributions over 3-momentum. By making use of the
 interpolating polynomials in lepton 3-momenta $p_{\mu}$($p_{e}$) [MeV]
we find the following  numerical fits to these distributions:
\begin{equation}
\begin{split}
\frac{d \bf{Br}_{K_{\mu 3\nu}}}{d p_{\mu}} =  &-2.893\times 10^{-22}+2.234 \times 10^{-22} p_{\mu}+2.043 \times 10^{-22} p_{\mu}^2+8.368 \times 10^{-26} p_{\mu}^3
\\ &-1.632 \times 10^{-26} p_{\mu}^4+7.338 \times 10^{-29} p_{\mu}^5-9.035
\times 10^{-32} p_{\mu}^6\,,
\end{split}
 \end{equation}
\begin{equation}
\begin{split}
\frac{d \bf{Br}_{K_{e 3\nu}}}{d p_{e}} =  &-1.180\times 10^{-21}+1.040 \times 10^{-20} p_{e}+1.990 \times 10^{-22} p_{e}^2-4.793 \times 10^{-25} p_{e}^3 \\ &-1.328 \times 10^{-26} p_{e}^4+6.474 \times 10^{-29} p_{e}^5-7.752 \times 10^{-32} p_{e}^6\,.
\end{split}
\end{equation}
The full branchings, obtained by integration of
interpolating polynomials differ from those obtained by integration of
exact ${\cal O}(p^2)$ results by less than one percent.

Integrating over the muon (electron)
momentum, one arrives at the following branching ratios:
\begin{displaymath}
\begin{array}{rclcrcl}
{\rm Br}(K^+ \rightarrow \mu^+ \nu_{\mu} \nu_{\mu} \bar \nu_{\mu} )
  &
= & 8.42 \times 10^{-18} & \hspace{1.5cm}
& {\rm Br}(K^+ \rightarrow e^+ \nu_{e} \nu_{\mu} \bar \nu_{\mu} )
& = & 2.00 \times 10^{-17}
\\ {\rm Br}(K^+ \rightarrow \mu^+ \nu_{\mu} \nu_{e} \bar \nu_{e} ) & = &
6.27\times 10^{-18} && {\rm Br}(K^+ \rightarrow e^+ \nu_{e} \nu_{e} \bar
\nu_{e} ) & = & 1.30 \times 10^{-22}
\\ {\rm Br}(K^+ \rightarrow \mu^+ \nu_{\mu} \nu_{\tau} \bar \nu_{\tau}
) &=&1.48\times 10^{-16} && {\rm Br}(K^+ \rightarrow e^+ \nu_{e} \nu_{\tau} \bar \nu_{\tau} ) &=& 2.38 \times 10^{-16}
\\ {\rm Br}(K^+ \rightarrow \mu^+ \nu_{\mu} \nu \bar \nu ) &=
&1.62\times 10^{-16} && {\rm Br}(K^+ \rightarrow e^+ \nu_{e} \nu \bar \nu ) &= &2.58 \times 10^{-16}
\end{array}
\end{displaymath}

Obtained branchings are expectingly small and are beyond foreseeable
direct registration (for instance NA$62$ at CERN is investigating the processes
of $6$ orders higher branchings). The question arises, how to
distinguish this weak process from others with similar observable
signature. The full answer is certainly beyond the scope of this
paper. An example of the irreducible background process is $K^+
\rightarrow \pi^+ \nu \bar \nu \rightarrow \mu^+ \nu_{\mu} \nu \bar
\nu$. Its contribution to the same final state as $K_{\mu3\nu}$ in the
finite volume detector can be
discriminated by increasing the pion boost, which correspondingly
suppress the probability of pion decay in flight. Moreover,
the muon spectra exhibits different
behavior, because the $\pi^+$ spectrum is growing throughout the whole
kinematically allowed region\,\cite{Komatsubara:1999te} (and therefore
$\mu^+$ spectrum as well) while in $K^+ \rightarrow \mu^+ \nu_{\mu}
\nu \bar \nu$ there is a clear maximum at $E_{\mu} \approx 155$
MeV. Thus although Br($K^+ \rightarrow \pi^+ \nu \bar \nu \rightarrow
\mu^+ \nu_{\mu} \nu \bar \nu$) $\approx 10^6 Br(K^+ \rightarrow \mu^+
\nu_{\mu} \nu \bar \nu) $, the study of low energy muons will give a
chance to distinguish the two processes.
\section{Conclusions}
\label{sec:conclusions}
To summarise, exploiting the ChPT we perform to the leading order in
momenta the calculation of lepton
spectra and partial widths of kaon decays $K_{\mu 3\nu}$ and $K_{e
  3\nu}$. The accuracy of all numerical integrations and numerical
fits to spectra are not worse than
 one percent. Recall, that the uncertainty associated with limiting
 the ChPT calculations by the
 leading order is significantly
worse, about few tens percent.

\vskip 0.3cm
We thank Yury Kudenko for suggesting to consider the problem of
kaon four-body leptonic decay and encouraging conversations at various
stages of the project. We thank Roman Lee and Artur Shaikhiev
for discussions on the phase space integration.  The work is supported
by the RSF grant 14-22-00161.
 \appendix

\section{Squared amplitudes of $K_{\mu 3\nu}$ and $K_{e 3 \nu}$}
\label{squared}
In this section we present the explicit forms of squared matrix
elements describing the four body decays $K_{\mu3\nu}$ and
$K_{e3\nu}$. To shorten the notations we denote the scalar product of
$4$-vectors $p_i$ and $p_j$ as $p_ip_j\equiv x_{ij}$
(considering $i<j$ for definiteness). Then the corresponding
squared amplitudes for decay $K_{\mu3\nu}$ into the final states
with electron and tau
neutrinos, $l=e,\tau$, are
  \begin{equation}
  \begin{split}
 {M}^2 =& \big|A \times \bar \nu_l (p_1) \gamma^{\mu} (1 - \gamma_5)
 \nu_{l} (p_2) \cdot\bar \nu_{\mu}(p_3) \gamma_{\mu} ( 1-\gamma_5 )
 \mu(p_4) + B \times \bar \nu_l (p_1) \hat p_0 (1 - \gamma_5) \nu_l
 (p_2) \\ \times & \bar \nu_{\mu} (p_3) (1 +
 \gamma_5)\mu(p_4) + C \times \bar \nu_l (p_1) \gamma_{\mu}
 (1-\gamma_5) \nu (p_2)\cdot\bar \nu_{\mu} (p_3)
 \gamma_{\mu}(1-\gamma_5) \hat p_0 \mu(p_4)\big|^2 =
\\
=& 256 A^2
 x_{13}x_{24} + 64 B^2 x_{34} \l 2x_{01}x_{02} - x_{12} M_K^2 \r + 256 C^2
 \l 2 x_{13}x_{02} x_{04} - M_K^2 x_{13} x_{24}\r
\\
-& 128 A B
 m_{\mu} \l x_{13}x_{02} +x_{01}x_{23} -x_{12}x_{03} \r - 512m_{\mu} A C
 x_{13}x_{02} + 128 B C \left[ 2 x_{02} \l x_{01}x_{34}
   \right.\right.
\\ +&\left.\left. x_{13}x_{04}
   - x_{03}x_{14} \r- M_K^2 \l x_{12}x_{34} +x_{13}x_{24}
   -x_{14}x_{23}\r \right]\,,
\end{split}
\end{equation}
where
\begin{equation}
\begin{split}
A &= 2 \l\sin^2 \theta+ \frac{M_K^2 -
  2x_{04}}{2  (M_K^2 - 2x_{04}+
  m^2_{\mu})}- \frac{M_K^2 - 2x_{01}}{ M_K^2 -
  2x_{01} - m_{l}^2}\right.
 \\ &+\left.\l \frac{1-2\sin^2\theta}{2}\r \l
 1+\frac{m_\mu^2}{2x_{12}+2x_{14}+2x_{24}}\r \r,\\
B &= -2 m_{\mu} \l \dfrac{1 - 2 \sin^2 \theta}{2x_{12} - 2x_{01} - 2x_{02}} -\dfrac{2\sin^2\theta}{2x_{12}+2x_{14}+2x_{24}}\r,\\
C &= - m_{\mu} \l \dfrac{1}{M_K^2
  -2x_{04} + m_{\mu}^2} +
  \dfrac{2\sin^2\theta}{2x_{12}+2x_{14}+2x_{24}} \r.
\end{split}
\end{equation}
Since $B$ and $C$ are proportional to mass of the charged lepton in
the final state, in case of the decay $K_{e3\nu}$ they can be
neglected. Then only one term in the squared matrix element survives,
that is proportional to $A^2$, where one must replace $m_\mu$ with
$m_e$.

Squared matrix elements for $K_{\mu3\nu}$ with only muon neutrinos in
the final set, $l=\mu$, read
\begin{equation}
\begin{split}
{M}^2 =& \big|A \times \bar \nu_l (p_1) \gamma^{\lambda} (1 - \gamma_5)
\nu_{l} (p_2) \cdot \bar \nu_{\mu}(p_3) \gamma_{\lambda} ( 1-\gamma_5 )
\mu(p_4) + B \times \bar \nu_l (p_1) \hat p_0 (1 - \gamma_5) \nu_l
(p_2) \\ \times & \bar \nu_{\mu} (p_3) (1 +
\gamma_5)\mu(p_4) + C \times \bar \nu_l (p_1) \gamma_{\lambda}
(1-\gamma_5) \nu (p_2) \cdot \bar \nu_{\mu} (p_3) \gamma^{\lambda}(1-\gamma_5)
\hat p_0 \mu(p_4) \\ +& D \times \bar \nu_l (p_3) \hat p_0 (1 -
\gamma_5) \nu_l (p_2)\cdot\bar \nu_{\mu} (p_1) (1 + \gamma_5)\mu(p_4) +
E \times \bar \nu_l (p_3) \gamma_{\lambda} (1-\gamma_5) \nu (p_2)
\\ \times & \bar \nu_{\mu} (p_1) \gamma^{\lambda}(1-\gamma_5) \hat p_0
\mu(p_4)\big|^2 = \\
=& 256 A^2 x_{13}x_{24} + 64 B^2 x_{34}
 \l 2x_{01}x_{02} - x_{12} M_K^2\r  + 256 C^2 x_{13} \l 2 x_{02}x_{04} -
M_K^2 x_{24}\r
\\
-& 128 A B m_{\mu} \l x_{13}x_{02}
  +x_{01}x_{23} -x_{12}x_{03} \r - 512m_{\mu} A C x_{13}x_{02} + 128 B
C \left[ 2 x_{02} \l x_{01}x_{34}  \right. \right.
\\
+& \left.\left. x_{13}x_{04} - x_{03}x_{14} \r-
  M_K^2 \l x_{12}x_{34} +x_{13}x_{24} -x_{14}x_{23} \r \right] + 64 D^2 x_{14} \l
2 x_{03} x_{02} - M_K^2 x_{23} \r
\\
+& 256 E^2 x_{13} \l 2 x_{02} x_{04}
- M_K^2 x_{24} \r - 128 m_{\mu} A D \l x_{12} x_{03} + x_{13}
  x_{02} - x_{01} x_{23}\r -512 m_{\mu} A E
\\
\times &  x_{13} x_{02} - 64 B D
\left[ 2x_{02} \l x_{01}x_{34} + x_{14}x_{03} - x_{13}x_{04}\r  -
  M_K^2 \l x_{12}x_{34} + x_{14}x_{23} - x_{13} x_{24}\r \right]
\\
+&
  128 B E \left[2
x_{02} \l x_{01} x_{34} + x_{13}x_{04} - x_{14}x_{03}\r - M_K^2
\l x_{12}x_{34} + x_{13}x_{24} - x_{14}x_{23}\r \right]
\\
+&    128 C D \left[ 2
x_{02}\l x_{03} x_{14} + x_{13}x_{04} - x_{34}x_{01}\r - M_K^2
\l x_{23}x_{14} + x_{13}x_{24} - x_{34}x_{12}\r \right]
\\
+&    512 C E x_{13} \l
  2x_{04}x_{02} - M_K^2 x_{24}\r  + 128 E D \left[2 x_{02} \l x_{03}x_{14} +
  x_{13}x_{04} - x_{34}x_{01}\r \right.
\\
-& \left. M_K^2 \l x_{23}x_{14} +  x_{13}x_{24} - x_{34}x_{12}\r \right]
\end{split}
\end{equation}
with
\begin{equation}
\begin{split}
A =& 2  \left[ \l 2\sin^2 \theta + \frac{M_K^2 -2 x_{04}}{
    M_K^2 - 2x_{04} +m^2_{\mu}} \r + \frac{ 1 - 2 \sin^2 \theta}{2}
\l 2+ \frac{m_{\mu}^2}{2x_{12} + 2x_{14} + 2x_{24}}\right.\right.
\\
+& \left. \left. \frac{m_{\mu}^2}{2x_{23} + 2x_{34} +2x_{24}} \r -
   \l \frac{M_K^2 - 2x_{01}}{M_K^2 - 2x_{01} -
    m_{\mu}^2} + \frac{M_K^2 - 2x_{03}}{M_K^2 - 2x_{03}- m_{\mu}^2} \r
\right]
\\
B =& 2  m_{\mu} \l \frac{2 \sin^2 \theta -
  1}{2x_{12} - 2x_{01} - 2x_{02}} + \frac{2 \sin^2 \theta}{2x_{12}
  +2x_{14} +2x_{24}}\r
\\
C =& - 2 m_{\mu}
\l\frac{1}{2(M_K^2 - 2x_{04} +m_{\mu}^2)} + \frac{\sin^2
  \theta}{2x_{12} + 2 x_{14} + 2 x_{24}}\r
\\
D=& - 2 m_{\mu} \l \frac{2 \sin^2 \theta - 1}{2x_{23} - 2x_{03} -
  2x_{02}} + \frac{2 \sin^2 \theta}{2x_{23} +2x_{34} +2x_{24}}\r
\\
E=& 2 m_{\mu} \l \frac{1}{2(M_K^2 - 2x_{04} +m_{\mu}^2)}
+ \frac{\sin^2 \theta}{2x_{23} + 2 x_{34} + 2x_{24}} \r
\end{split}
\end{equation}
Again, in case of the decay $K_{e3\nu}$ all the coefficients above
except $A$ and $E$ are proportional to mass of the charged lepton in
the final state. Thus only the single term
proportional to $A^2$ saturates the squared matrix element of
$K_{e3\nu}$, and $m_e$ replaces $m_\mu$ in the expressions of $A$
presented above.

\section{Integration over the phase space of $K_{\mu 3\nu}$ and
  $K_{e 3\nu}$}
\label{phase-space}
  We examine the four particle decay and phase density of final states
has relatively complicated from. The main idea is to examine a chain
of two particle decays from the rest frame of kaon. Thus we write all
momenta in the fixed frame from the very beginning. As a consequence
of neutrinos being massless, momenta of decay products are unambiguous
and resulting formula has a rather simple form.

In the following section we are using notation $p_i=(E_i, \vec p_i)$
for 4-vector $p_i$ and ${\bf p}$ for length of 3-vector $\vec
p$. Formula of 2-particle phase space for decay $q \rightarrow q_1 +
q_2$ reads\,\cite{Byckling:1971vca}
\begin{equation}
\label{R2}
R_2(q,q_1,q_2) = \frac{1}{4} \int d \Omega_1 \sum_{a=+,-} (\textbf{q}_1^{a})^2(E \textbf{q}_1^{a} - \textbf{q} E_1^{a} \cos \theta)^{-1}\,,
\end{equation}
where $\textbf{q}_1^{+,-}$ are solutions to the following equation,
\[
E - E_1 - (\textbf{q}^2 + \textbf{q}_1^2 - 2 \textbf{q q}_1
\cos\theta_1 + m_2^2)^{1/2} =0\,,
\]
so that $\cos \theta_1$ is angle between $\vec q$ and $\vec q_1$, $m_2^2 =
q_2^2$. The r.h.s. of eq.\,\eqref{R2}
contains Lorentz-non-invariant variables, but
after integration $R_2$ still depends only on $q^2$ as it must
be. Nevertheless we will denote it as $R_2(q,q_1,q_2)$ to specify
particles, involved in decay.

Decomposition of n-particle phase
space $R_{n}$ is performed with the following formula\,\cite{Byckling:1971vca}
\begin{equation}
\label{Rn}
R_{n}(M_n^2) = \int_{\mu_{n-1}^2}^{(M_n-m_n)^2} d M_{n-1}^2 R_2(k_n,k_{n-1},p_n)R_{n-1}(M_{n-1}^2) \,,
\end{equation}
where $m_i = p_i^2$, $\mu_i \equiv \sum_1^i m_i$, $k_i \equiv \sum_1^i
p_i$, $M_n^2 = k_n^2$. Thus, applying  the reduction formula
\eqref{Rn} twice, substituting common $R_2$ expressions and
integrating out $3$ trivial angles, we arrive at the following master
formula

\begin{equation}
\begin{split}
R_4(M_K^2)  = \frac{\pi ^2}{8} \int^{(M_K - m_{4})^2}_{0} d M^2_3
\int^{M_3^2}_{0} d M_2^2 \int_{-1}^{1} d \cos \theta_1 \int_{-1}^1 d \cos \theta_2
\int_0^{2 \pi} d \phi \\ \times \frac{[(M_K^2 - (M_3 +
    m_{4})^2)(M_K^2 -(M_3 - m_{4})^2)]^{1/2}\times \textbf{p}_3^2
  \times \textbf{p}_2^2}{2 M_K^2 \times (E_{123} \textbf{p}_3 -
  \textbf{p}_{123} E_3 \cos \theta_1) \times (E_{12} \textbf{p}_2 -
  \textbf{p}_{12} E_2 \cos \theta_2)}\,.
\end{split}
\end{equation}

\begin{figure}[!htb]
   \center{\includegraphics[scale=1]{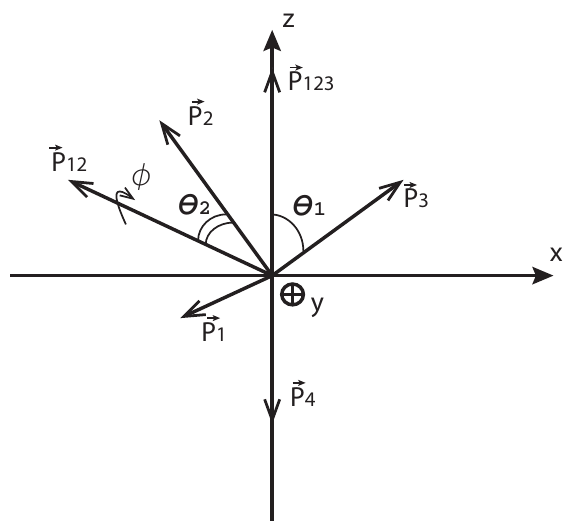}}
	 \caption{\label{vectors} Definition of integration variables
(2D sketch).}
\end{figure}

Here we adopt the useful auxiliary momentum variables
$p_{12}=(E_{12},\vec p_{12}) = p_1 + p_2$; $p_{123} = (E_{123},\vec
p_{123} ) = p_1 + p_2 + p_3$, so that $M^2_2 = p_{12}^2$, $M^2_3 =
p_{123}^2$, and angle variables
$\theta_1 \equiv \angle (\vec p_3 \,, \vec p_{123})$,
$\theta_2 \equiv \angle (\vec p_2,\vec p_{12})$,  $\phi$ is
rotation angle of plane $(\vec p_{12} \,, \vec p_2)$ around $\vec
p_{12}$, as demonstrated in Fig.\,\ref{vectors}.
The decay rates of $K_{l 3\nu}$ are found by integrating the
corresponding squared amplitudes \eqref{amplitude} over the
above defined phase space,
\begin{equation}
\Gamma_{K_{l3\nu}} = \int \frac{1}{2M_K (2 \pi)^8} \left| {\cal M}
\right|^2 d R_4\,,
\end{equation}
in the cases with two identical
neutrinos in final state we must also divide the obtained in this way
result by $2$.
Now substituting $M_3^3 = (p_0 - p_4)^2$
it is straightforward to obtain the differential spectra:
\begin{equation}
\begin{split}
\frac{d\Gamma_{K_{l3\nu}}}{d{\bf p_{4}}}  = \frac{1}{(2 \pi)^6} \int^{M_3^2}_{0} d M_2^2 \int_{-1}^{1} d \cos \theta_1 \int_{-1}^{1} d \cos \theta_2 \int_{0}^{2 \pi} d \phi \\ \times \frac{[(M_K^2 - (M_3 + m_{4})^2)(M_K^2 -(M_3 - m_{4})^2)]^{1/2}\times \textbf{p}_3^2 \times \textbf{p}_2^2}{2 M_K^2 \times (E_{123} \textbf{p}_3 - \textbf{p}_{123} E_3 \cos \theta_1) \times (E_{12} \textbf{p}_2 - \textbf{p}_{12} E_2 \cos \theta_2)} \times \frac{\textbf{p}_4  \left| {\cal M} \right|^2}{32 E_4} \,,
\end{split}
\end{equation}
where $M_3^2 = M_K^2+m_{4}^2 - 2 M_K E_{4}$.
Then the branching fraction is defined as
\[
\text{Br}_{K_{l3\nu}}\equiv \tau_K \cdot \Gamma_{K_{l3\nu}}\,,
\]
where $\tau_K$ is the kaon lifetime. For the numerical estimates we
substitute the kaon decay constant $F_K\approx 111$\,MeV
\cite{Agashe:2014kda} for the
meson decay constant $F$.
\bibliographystyle{apsrev4-1}
%
\bibliography{refs}

\end{document}